\documentclass[conference]{IEEEtran}
\IEEEoverridecommandlockouts

\usepackage{cite}
\usepackage{amsmath,amssymb,amsfonts}
\usepackage{algorithmic}
\usepackage{graphicx}
\usepackage{textcomp}
\usepackage{xcolor}
\usepackage{xspace}
\usepackage{makecell}
\usepackage{multirow}
\usepackage{booktabs}
\usepackage{array}
\usepackage{subcaption}
\usepackage{hyperref}
\usepackage{cleveref}
\Crefformat{figure}{#2Fig.~#1#3}
\Crefmultiformat{figure}{Figs.~#2#1#3}{ and~#2#1#3}{, #2#1#3}{ and~#2#1#3}

\newcommand{\name}{Lachesis\xspace}

\def\BibTeX{{\rm B\kern-.05em{\sc i\kern-.025em b}\kern-.08em
    T\kern-.1667em\lower.7ex\hbox{E}\kern-.125emX}}
\begin{document}

\title{\name: Predicting LLM Inference Accuracy using Structural Properties of Reasoning Paths}

\author{\IEEEauthorblockN{Naryeong Kim}
\IEEEauthorblockA{\textit{School of Computing} \\
\textit{KAIST}\\
Daejeon, Republic of Korea \\
naryeong.kim@kaist.ac.kr}
\and\IEEEauthorblockN{Sungmin Kang}
\IEEEauthorblockA{\textit{School of Computing} \\
\textit{KAIST}\\
Daejeon, Republic of Korea \\
sungmin.kang@kaist.ac.kr}
\and
\IEEEauthorblockN{Gabin An}
\IEEEauthorblockA{\textit{School of Computing} \\
\textit{KAIST}\\
Daejeon, Republic of Korea \\
gabin.an@kaist.ac.kr}
\and
\IEEEauthorblockN{Shin Yoo}
\IEEEauthorblockA{\textit{School of Computing} \\
\textit{KAIST}\\
Daejeon, Republic of Korea \\
shin.yoo@kaist.ac.kr}
}

\maketitle

\begin{abstract}
Large Language Models are increasingly used to build agents to perform more complex tasks. As LLMs perform more complicated reasoning through longer interactions, self-consistency, i.e., the idea that the answer obtained from sampling and marginalising a number of multiple independent inferences is more likely to be correct, has received much attention as a simple validation technique. This paper aims to empirically verify this intuitive hypothesis by predicting the correctness of answers obtained using self-consistency from properties of the samples of reasoning paths. We introduce \name, a predictive model for self-consistency based LLM inferences, and empirically evaluate it using AutoFL, a recently proposed LLM-based fault localisation technique, as the target technique that uses self-consistency. \name converts collected reasoning paths from AutoFL using specifically designed reasoning path representations, and trains LSTM and GCN models to predict whether a given set of reasoning paths would result in a correct answer. The results suggest that \name can predict the correctness of answers with a precision of up to 0.8136, highlighting the possibility of training a predictive model that can allow early termination of inferences that are not likely to be successful. 
\end{abstract}


\section{Introduction}
\label{sec:introduction}

Large Language Models (LLMs) are rapidly being adopted to automate various stages of the software development lifecycle due to their capability to perform semantic reasoning across the barrier between natural and programming language~\cite{Fan2023yu}. While the chatbot model, where an LLM after instruct-tuning converses with a human user, has been the initial mode of interaction, increasingly LLMs are being used to build more complicated agents~\cite{Feldt2023ax} for tasks like program repair~\cite{Kang2023ad,Yang2024aa}, testing~\cite{Yoon2024aa}, and fault localisation~\cite{kang2024quantitative}. These agents are assisted by various in-context learning techniques, such as Chain-of-Thoughts~\cite{wei2022chain}, self-consistency~\cite{wang2022self}, and ReAct~\cite{yao2022react}.

Self-consistency~\cite{wang2022self} in particular has received much attention because of the simplicity of the technique: instead of greedily decoding LLM outputs, self-consistency posits that sampling and marginalising reasoning paths can produce more accurate answers. That is, if multiple inferences sampled independently reaches the same answer, the answer is more likely to be correct. Since the only required computation is multiple sampling of LLMs and the majority voting, self-consistency has been widely studied and adopted~\cite{Ahmed2023aa,kang2024quantitative}. 



Despite the success of self-consistency, there are corresponding cost concerns when using it as well, as to use self-consistency one must query an LLM multiple times, which is computationally costly (and as an extension, potentially harmful for the environment~\cite{Strubell2019aa}). To find a clue to overcome this challenge, we look at the hypothesis that self-consistency relies on: for logical problems, there tend to be multiple reasoning paths that lead to the correct answer~\cite{wang2022self}. This is interestingly reminiscent of fitness landscape analyses, where an optimum can be said to have a broader `basins of attraction', in that more random solutions will end up at the optima~\cite{Langdon2002aa}. In optimization, one may imagine multiple solutions `aggregating' at the neighbours of an optimum before actually arriving at the optimum. Similarly, with self-consistency, one may observe a convergence of intermediate reasoning steps \emph{before} observing convergence towards the final answer, and use that to make an early prediction of whether the LLM will get the answer correct for this question. Simply put, can we predict the result of self-consistency before LLM inference generates the answer? This may enable early termination of likely unsuccessful LLM inferences, thereby reducing the cost of self-consistency. 

As a preliminary study, this paper empirically investigates whether such predictions are feasible by looking at the traces of multiple inferences. We introduce \name, a predictive model that aims to classify whether a given set of LLM reasoning paths will result in a correct answer. \name is evaluated in conjunction with an LLM-based fault localization tool, AutoFL~\cite{kang2024quantitative}, as the target, and can predict the success of inferences with precision of up to 0.8136.

\section{Methodology}
\label{sec:methodology}

\subsection{AutoFL \& Self-Consistency}
\label{sec:autofl}

AutoFL~\cite{kang2024quantitative} is a recently proposed method-level Fault Localization (FL) technique that leverages LLMs. FL requires extensive contextual information, such as contents of source code and test coverage, which are often too long to fit in the context window of LLMs. To overcome this issue, AutoFL equips the LLM with a set of functions (tools), enabling the LLM to autonomously gather relevant information, instead of receiving all contextual details upfront, \textit{ReAct}~\cite{yao2022react}. Given the initial failure information, AutoFL uses the set of given functions to collect details on class- and method-level coverage of failing test cases, method snippets, and documentation. AutoFL is given $N$ such function calls to determine the buggy method as its final answer: it repeats this process $R$ times for a single bug and aggregates to generate a final list of suspicious methods to present to the user. While AutoFL is an ideal candidate for \name, note that \name can generalise to any LLM-based agents that uses a set of functions (i.e., tools) as reasoning steps.


\subsection{Representation of Reasoning Paths}
\label{sec:representation}

Wang et al.~\cite{wang2022self} hypothesise that complex reasoning tasks for LLMs would typically allow multiple paths that reach the correct solution. In turn, we posit that, if the hypothesis about reasoning paths is true, it would also be possible to predict the correctness of the reasoning based on the shape of the combined multiple reasoning paths. 
\name uses two representations of reasoning paths in AutoFL: LLM Inference Matrix, and LLM Inference Graphs.

\subsubsection{LLM Inference Matrix}
\label{sec:lim}

\begin{figure}[t]
    \centering
    \includegraphics[width=1\linewidth]{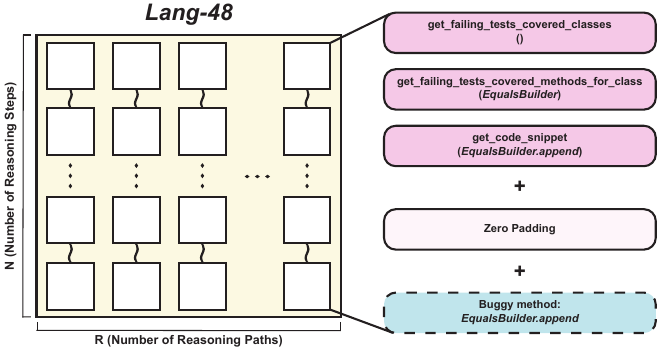}
    \caption{Structure of LIM}
    \label{fig1_LIM}
\end{figure}

An LLM Inference Matrix (LIM) is a collection of reasoning paths: each column of the matrix corresponds to a single reasoning path. Consequently, with AutoFL inference data, the corresponding LIM would have $R$ columns, each of which consists of up to $N$ embedding vectors (see Section~\ref{sec:embedding}) that occupy different rows, each corresponding to a function call, as shown in \Cref{fig1_LIM}. If AutoFL returns an answer before completing all $N$ function calls, the remaining positions in the chain are padded with zeros.

\subsubsection{LLM Inference Graph}
\label{sec:lig}

\begin{figure}[t]
    \centering
    \includegraphics[width=0.9\linewidth]{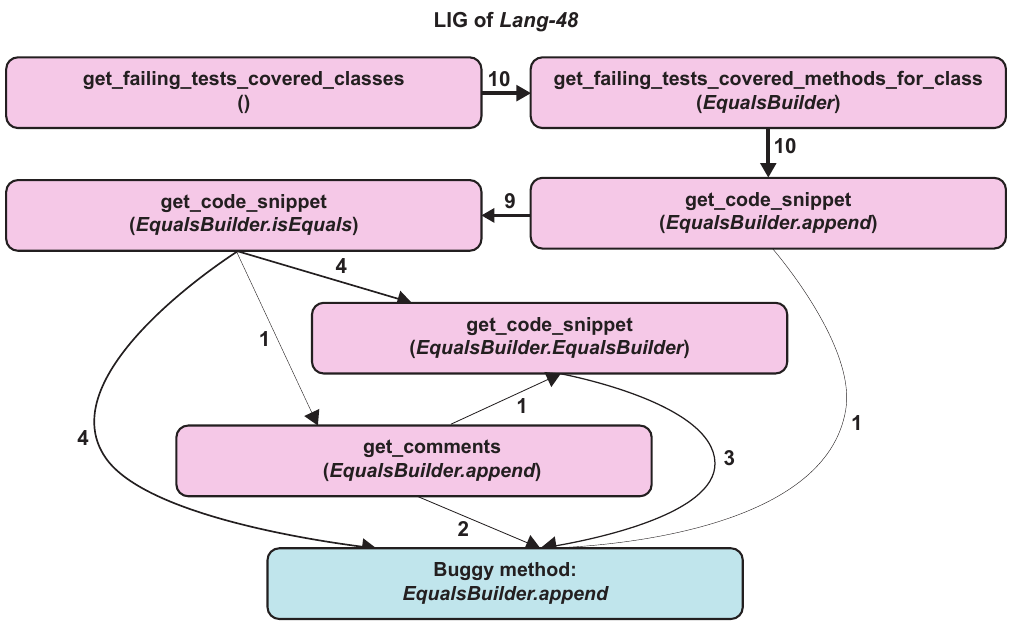}
    \caption{Example graph of LIG: Edge weight is represented by the width of arrows.}
    \label{fig2_LIG}
\end{figure}

An LLM Inference Graph is a digraph representation of multiple reasoning paths: each node is a specific reasoning step (a specific function call, or FL outcome, in the case of AutoFL), whereas each edge connects two consecutive reasoning steps (i.e., nodes) in the order in which they were performed. Edges are weighted to represent multiple reasoning paths with the same subsequence of reasoning actions. Ans example is given in \Cref{fig2_LIG}.

\subsection{Embedding Reasoning Steps}
\label{sec:embedding}

\begin{figure} [t]
    \centering
    \includegraphics[width=0.8\linewidth]{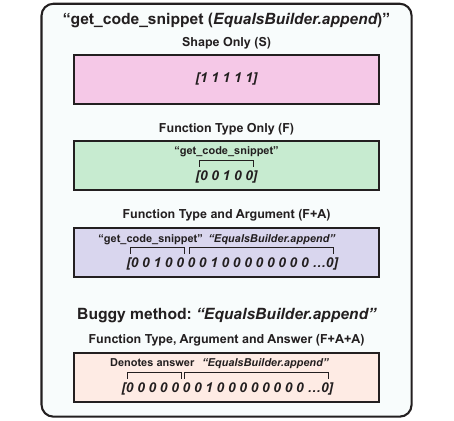}
    \caption{Different Embeddings of Reasoning Steps and Final Answers, taken from Lang-48}
    \label{fig3_embedding}
\end{figure}

A reasoning step of AutoFL is either a specific function call, or an FL result. We experiment with different information contents that go into the representation of a reasoning step. Embedding examples are provided in \Cref{fig3_embedding}.

\subsubsection{Shape Only (S)} The bare minimum information that can support the self-consistency hypothesis is the shape of reasoning paths, i.e., whether multiple paths converge to a single answer that is correct. To verify this, Shape Only (S) representation uses a vector filled with ones of length 5 as the node embedding for both function call nodes and answer nodes, without any additional information. This helps us to evaluate how accurate the prediction can be when using only the shape of the LIG. Unlike other embeddings of reasoning steps, Shape Only can only be used with LIG.

\subsubsection{Function Type Only (F)} Kang et al.~\cite{kang2024quantitative} reports a pattern in AutoFL function calls: the LLM gradually narrows down its search scope by looking at code snippets and documentations. To reflect this, Function Only (F) representation uses one-hot vectors to represent different function types used at each reasoning step. Since AutoFL provides three common functions and one unique function each for BugsInPy and Defects4J, we use a one-hot vector of length five to represent the function type used in a reasoning step.

\subsubsection{Function Type and Arguments (F+A)} In addition to Function Type representation, this representation also includes the specific function argument values used with function calls. We hope to capture the cases where AutoFL is gradually narrowing down the candidate fault locations. For example, if multiple reasoning paths look at code snippet and documentation of the same method, it can be a sign that the method is more likely to be the correct location of the fault. For F+A, we first collect all classes or methods that appeared during reasoning for each bug, both as function arguments and as AutoFL answers, excluding those that failed to interact with AutoFL because they do not exist in the repository. These are represented as one-hot vectors with one extra position beyond the collection size, with the last position reserved for those arguments that are not included in the collection due to the non-existence. These vectors are then concatenated with the function type one-hot vectors.

\subsubsection{Function Type, Arguments, and Answer (F+A+A)} In addition to Function Type and Argument, this representation includes the final answer given by AutoFL. Similarly to F+A, we collect all methods and classes that have appeared in reasoning paths. Since AutoFL can provide multiple answers as buggy method, we mark all corresponding position in the vector with ones to represent the answers. The vector is then appended to the path of the function calls.

\subsection{Prediction Model}
\label{sec:model}

Once \name represents reasoning paths in either LIM or LIG, it predicts whether the given set of reasoning paths contains the correct answer. In the case of AutoFL, we define a set of reasoning paths to contain a correct answer if AutoFL ranks the faulty method at the top after it computes the ranking score by voting~\cite{kang2024quantitative}. Consequently, we use the AutoFL results to label LIMs and LIGs. \name uses an Long Short Term Memory (LSTM)~\cite{Hochreiter1997ak} model to predict the correctness of LIMs, and a Graph Convolution Network (GCN)~\cite{Kipf2016rb} model to predict the correctness of LIGs.

\subsubsection{LSTM}

The LSTM architecture consists of multiple LSTM layers and a single fully connected layer. When passing through the LSTM layer, multiple reasoning paths are stacked together, so that the LSTM can process all paths step-wise along the time axis in the order of function calls (i.e., the LSTM process $R$ first reasoning steps, followed by $R$ second reasoning steps, up to $N$-th reasoning steps).

\subsubsection{GCN}

In the GCN model, the data passes a sequence of GCN layer, ReLU activation function, and dropout layer multiple times, allowing the model to learn the graph structures effectively. Subsequently, the final GCN layer, followed by global mean pooling and a fully connected layer, performs the final classification.  

\section{Evaluations}
\label{sec:evaluations}

\subsection{Experimental setups}
\label{sec:experimental_setup}

\subsubsection{Dataset Construction} 

In the original experiments of AutoFL~\cite{kang2024quantitative}, $R$ and $N$ were set to 5 and 10, respectively, that is, for each bug, AutoFL performs 5 inferences, each of which can have up to 10 function calls. AutoFL was evaluated using 798 bugs from the BugsInPy~\cite{widyasari2020bugsinpy} and Defects4J~\cite{just2014defects4j} datasets.\footnote{Our code and data are available at \url{https://figshare.com/s/d7fc515a52a379ae47ea}.}

For this study, we configure AutoFL to run 10 times per bug (i.e., $R=10$), and create a dataset from these 10 reasoning paths, and keep $N$ as 10. To simplify data preparation and model training, we limited our dataset to buggy program versions containing only one ground truth buggy method from the original AutoFL evaluation dataset. Bugs from Closure of Defects4J were excluded due to the high cost of LLM executions. This results in a dataset of 456 bugs, 294 from BugsInPy and 162 from Defects4J. AutoFL correctly ranks the buggy methods at the top for 307 of these bugs.

\subsubsection{Baselines}
We compare the results with AutoFL confidence scores~\cite{kang2024quantitative}. Each of $R$ inferences conducted by AutoFL produces answers (i.e., candidate faulty locations), to which AutoFL votes. The voting produces final scores for candidate faulty methods, and the method with the most votes is chosen as the final answer. The highest score then becomes the confidence score, serving as a measure of confidence in the final answer. The formula used to measure the confidence score is as follows: \begin{equation} \text{confidence} = \max_{m \in M} \text{score} (m) \end{equation} where $M$ denotes the set of methods covered by failing tests and $\text{score}(m)$ denotes the voting score of each method. The AutoFL-confidence metric requires the final answer of the reasoning process to measure accuracy, as demonstrated here. Using this confidence score, we calculated the classification accuracy, ROC-AUC, precision, and recall, to compare with our approach. In addition, we report baseline scores by assuming all prediction results are correct to ensure that \name can handle the imbalance in the dataset.

\subsubsection{Metrics}

We use the standard evaluation metrics for classification: accuracy, ROC-AUC, precision, and recall. With AutoFL-Confidence, we set the threshold to $0.5$, which means that the final result of AutoFL is predicted to be correct (i.e., the actual faulty method will be ranked at the top) when the confidence score is greater than or equal to $0.5$.

\subsubsection{Hyperparameter Tuning}

\begin{table}[t]
\centering
\caption{Hyperparameter Settings\label{tab:hyperparameters}}
\scalebox{0.8}{
\begin{tabular}{l|rrr|rrrr}
\toprule
            & \multicolumn{3}{c|}{LSTM} & \multicolumn{4}{c}{GCN} \\
            & F  & F+A & F+A+A & S   & F   & F+A & F+A+A \\ \midrule
Layers      & 1  & 2   & 1     & 3   & 3   & 3   & 3     \\
Hidden dim. & 32 & 128 & 32    & 128 & 64  & 64  & 64    \\  
Batch       & 64 & 16  & 32    & 32  & 16  & 16  & 16    \\
Dropout     & 0  & 0.5 & 0     & 0.3 & 0.8 & 0.8 & 0.8   \\
\bottomrule
\end{tabular}}
\end{table}


The number of LSTM layers, the number of iterations of layer sequences including GCN layers, hidden dimensions, batch size and dropout probability are manually fine-tuned. The tuned hyperparameters are presented in the \Cref{tab:hyperparameters}. A 10-fold cross-validation is employed to mitigate the risk of over-fitting. Both LSTM and GCN models are trained for 50 and 100 epochs each with a learning rate of 0.001. We report test accuracy, ROC-AUC, precision, and recall from the epoch with the highest test accuracy.

\begin{figure}[t]
    \centering
    \begin{subfigure}[b]{0.15\textwidth}
        \centering
        \includegraphics[width=\textwidth]{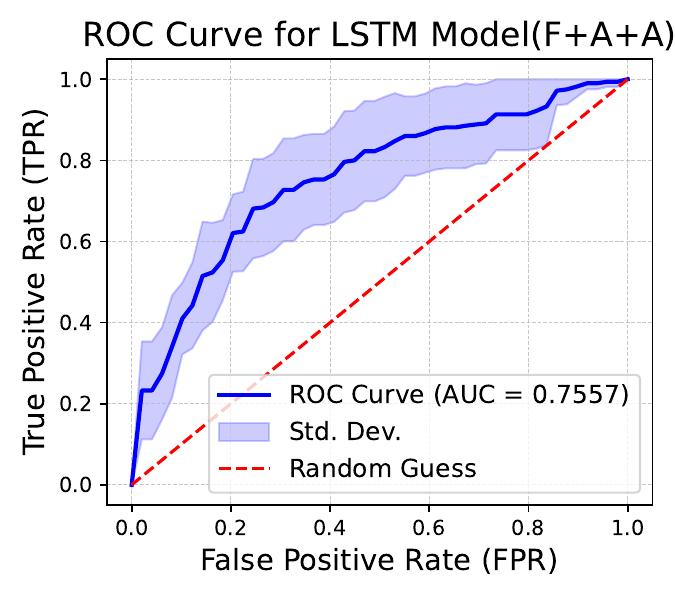}
        \caption{LSTM}
        \label{fig4_lstm_roc}
    \end{subfigure}
    \begin{subfigure}[b]{0.15\textwidth}
        \centering
        \includegraphics[width=\textwidth]{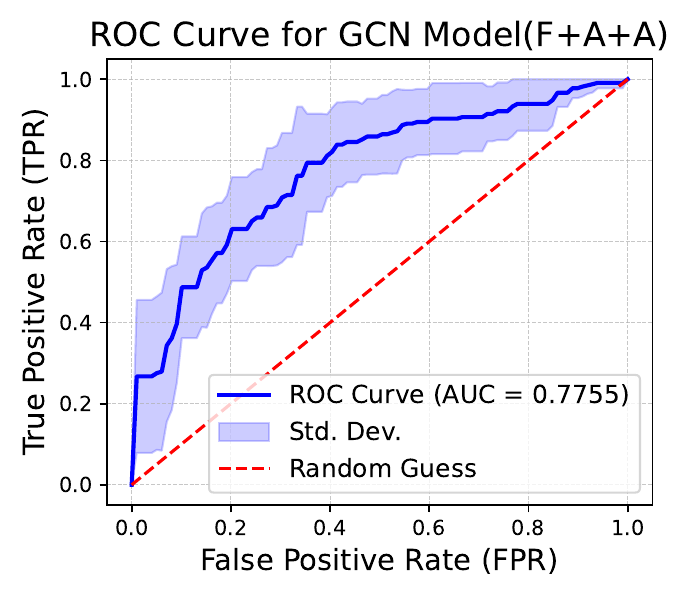}
        \caption{GCN}
        \label{fig5_gcn_roc}
    \end{subfigure}
    \begin{subfigure}[b]{0.15\textwidth}
        \centering
        \includegraphics[width=\textwidth]{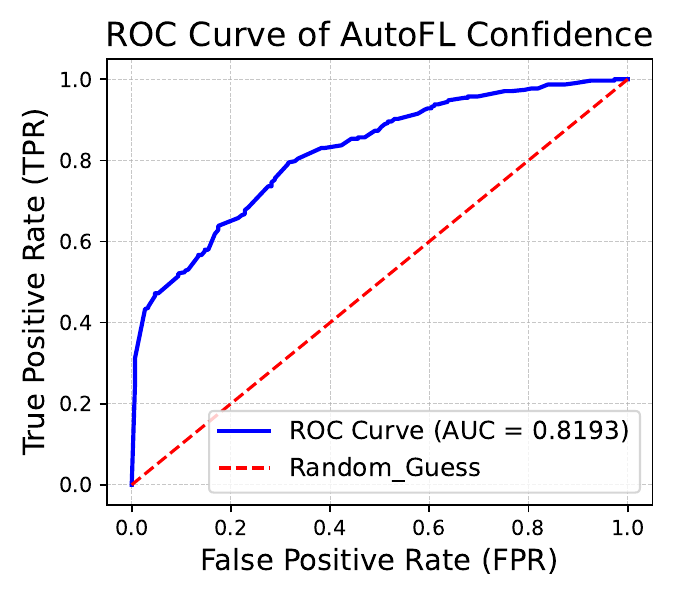}
        \caption{AutoFL-S.C.}
        \label{fig6_autofl_roc}
    \end{subfigure}
    \caption{ROC-AUC Graphs}
    \label{fig7_roc_graphs}
\end{figure}

\subsection{Results}
\label{sec:results}
    
As shown in \Cref{tab:results}, \name achieves performance comparable to AutoFL confidence, although AutoFL confidence slightly outperforms it. A similar trend is observed in ROC-AUC plots in \Cref{fig7_roc_graphs}: the average ROC-AUC from $k$-fold evaluation of \name is slightly outperformed by AutoFL confidence score. However, \name achieves competitive precision (GCN, F+A+A) and the highest recall (GCN, S). We argue that the high precision would be better in the expected use case, as it would also lead to fewer false positives.  

\begin{table}[t]
\centering
\caption{Performance of Prediction Models and Baselines\label{tab:results}}
\scalebox{0.8}{
\begin{tabular}{llrrrr}
\toprule
\multicolumn{2}{l}{Method}        & Accuracy        & ROC-AUC         & Precision       & Recall \\ \midrule
\multirow{3}{*}{LSTM}  & F        & 0.7063          & 0.7356          & 0.7570          & 0.8302 \\
                       & F+A      & 0.7149          & 0.6870          & 0.7662          & 0.8268 \\
                       & F+A+A    & 0.7191          & 0.7557          & 0.7711          & 0.8272 \\ \midrule
\multirow{3}{*}{GCN}   & S        & 0.6900          & 0.7791          & 0.7323          & \textbf{0.9182} \\
                       & F        & 0.7235          & 0.7866          & 0.7751          & 0.8524 \\
                       & F+A      & 0.7454          & 0.7723          & 0.8022          & 0.8332 \\
                       & F+A+A    & 0.7454          & 0.7755          & 0.8136          & 0.8172 \\ \midrule
\multicolumn{2}{l}{AutoFL Conf.}  & \textbf{0.7610} & \textbf{0.8193} & \textbf{0.8173} & 0.8306 \\
\multicolumn{2}{l}{Baseline}      & 0.6732          & -               & 0.6732          & 1.0000 \\
\bottomrule
\end{tabular}}
\end{table}

We note that both the LSTM and GCN models show a tendency to perform better as more information about arguments and answers are provided. This may be due to the fact that the additional information can reveal how AutoFL gradually narrows down the location of the faults by making function calls to a specific location. In addition, the GCN, which provides a more intuitive representation of the structure between function calls, outperforms the LSTM overall. Thus, the integration of function call information with structural information seems to have a synergistic effect. 

On the other hand, AutoFL-confidence is not designed to provide scores specifically for 0.5-threshold binary classification, so its calculated classification accuracy, precision, and recall scores may not serve as fully comparable metrics. As described before (Section~\ref{sec:autofl}), computing AutoFL confidence score requires the inference run to finish itself, whereas 
while configurations of \name such as F and F+A can perform well by leveraging limited information. This, in turn, suggests that predictions based on partial data may be possible, leading to the possibility of early terminations.

\section{Conclusion}

We present \name, a predictive model that aims to classify sets of LLM reasoning paths based on whether they will result in correct answers or not. \name is based on the hypothesis behind self-consistency, i.e., there tend to be multiple reasoning paths that lead to the correct answer. This allows \name to predict the correctness of the final answer based on the structural properties of the reasoning paths. \name achieves precision of 0.8136 when applied to reasoning of AutoFL, an LLM-based Fault Localisation technique. Future work will investigate whether \name can be extended to enable early termination of LLM inferences that are unlikely to be correct.

\bibliographystyle{IEEEtran}
\bibliography{ref}

\begin{thebibliography}{10}
\providecommand{\url}[1]{#1}
\csname url@samestyle\endcsname
\providecommand{\newblock}{\relax}
\providecommand{\bibinfo}[2]{#2}
\providecommand{\BIBentrySTDinterwordspacing}{\spaceskip=0pt\relax}
\providecommand{\BIBentryALTinterwordstretchfactor}{4}
\providecommand{\BIBentryALTinterwordspacing}{\spaceskip=\fontdimen2\font plus
\BIBentryALTinterwordstretchfactor\fontdimen3\font minus
  \fontdimen4\font\relax}
\providecommand{\BIBforeignlanguage}[2]{{%
\expandafter\ifx\csname l@#1\endcsname\relax
\typeout{** WARNING: IEEEtran.bst: No hyphenation pattern has been}%
\typeout{** loaded for the language `#1'. Using the pattern for}%
\typeout{** the default language instead.}%
\else
\language=\csname l@#1\endcsname
\fi
#2}}
\providecommand{\BIBdecl}{\relax}
\BIBdecl

\bibitem{Fan2023yu}
A.~Fan, B.~Gokkaya, M.~Harman, M.~Lyubarskiy, S.~Sengupta, S.~Yoo, and J.~M.
  Zhang, ``Large language models for software engineering: Survey and open
  problems,'' in \emph{Proceedings of the 45th IEEE/ACM International
  Conference on Software Engineering: Future of Software Engineering}, ser.
  ICSE-FoSE.\hskip 1em plus 0.5em minus 0.4em\relax IEEE, May 2023, pp. 31--53.

\bibitem{Feldt2023ax}
R.~Feldt, S.~Kang, J.~Yoon, and S.~Yoo, ``Towards autonomous testing agents via
  conversational large language models,'' in \emph{Proceedings of the 38th
  IEEE/ACM International Conference on Automated Software Engineering (ASE)},
  ser. ASE 2023, 2023, pp. 1688--1693.

\bibitem{Kang2023ad}
S.~Kang, B.~Chen, S.~Yoo, and J.-G. Lou, ``Explainable automated debugging via
  large language model-driven scientific debugging,'' \emph{arXiv preprint
  arXiv:2304.02195}, 2023.

\bibitem{Yang2024aa}
J.~Yang, C.~E. Jimenez, A.~Wettig, K.~Lieret, S.~Yao, K.~Narasimhan, and
  O.~Press, ``Swe-agent: Agent-computer interfaces enable automated software
  engineering,'' 2024.

\bibitem{Yoon2024aa}
J.~Yoon, R.~Feldt, and S.~Yoo, ``Intent-driven mobile gui testing with
  autonomous large language model agents,'' in \emph{Proceedings of the 16th
  IEEE International Conference on Software Testing, Verification and
  Validation}, ser. ICST 2024, 2024, pp. 129--139.

\bibitem{kang2024quantitative}
S.~Kang, G.~An, and S.~Yoo, ``A quantitative and qualitative evaluation of
  llm-based explainable fault localization,'' \emph{Proceedings of the ACM on
  Software Engineering}, vol.~1, no. FSE, pp. 1424--1446, 2024.

\bibitem{wei2022chain}
J.~Wei, X.~Wang, D.~Schuurmans, M.~Bosma, F.~Xia, E.~Chi, Q.~V. Le, D.~Zhou
  \emph{et~al.}, ``Chain-of-thought prompting elicits reasoning in large
  language models,'' \emph{Advances in neural information processing systems},
  vol.~35, pp. 24\,824--24\,837, 2022.

\bibitem{wang2022self}
X.~Wang, J.~Wei, D.~Schuurmans, Q.~Le, E.~Chi, S.~Narang, A.~Chowdhery, and
  D.~Zhou, ``Self-consistency improves chain of thought reasoning in language
  models,'' \emph{arXiv preprint arXiv:2203.11171}, 2022.

\bibitem{yao2022react}
S.~Yao, J.~Zhao, D.~Yu, N.~Du, I.~Shafran, K.~Narasimhan, and Y.~Cao, ``React:
  Synergizing reasoning and acting in language models,'' \emph{arXiv preprint
  arXiv:2210.03629}, 2022.

\bibitem{Ahmed2023aa}
T.~Ahmed and P.~Devanbu, ``Better patching using llm prompting, via
  self-consistency,'' in \emph{2023 38th IEEE/ACM International Conference on
  Automated Software Engineering (ASE)}, 2023, pp. 1742--1746.

\bibitem{Strubell2019aa}
E.~Strubell, A.~Ganesh, and A.~McCallum, ``Energy and policy considerations for
  deep learning in {NLP},'' in \emph{Proceedings of the 57th Annual Meeting of
  the Association for Computational Linguistics}, A.~Korhonen, D.~Traum, and
  L.~M{\`a}rquez, Eds.\hskip 1em plus 0.5em minus 0.4em\relax Florence, Italy:
  Association for Computational Linguistics, Jul. 2019, pp. 3645--3650.

\bibitem{Langdon2002aa}
W.~B. Langdon and R.~Poli, \emph{Foundations of Genetic Programming}.\hskip 1em
  plus 0.5em minus 0.4em\relax Springer-Verlag, 2002.

\bibitem{Hochreiter1997ak}
S.~Hochreiter and J.~Schmidhuber, ``Long short-term memory,'' \emph{Neural
  Computation}, vol.~9, no.~8, pp. 1735--1780, 11 1997.

\bibitem{Kipf2016rb}
T.~N. Kipf and M.~Welling, ``Semi-supervised classification with graph
  convolutional networks,'' \emph{CoRR}, vol. abs/1609.02907, 2016.

\bibitem{widyasari2020bugsinpy}
R.~Widyasari, S.~Q. Sim, C.~Lok, H.~Qi, J.~Phan, Q.~Tay, C.~Tan, F.~Wee, J.~E.
  Tan, Y.~Yieh \emph{et~al.}, ``Bugsinpy: a database of existing bugs in python
  programs to enable controlled testing and debugging studies,'' in
  \emph{Proceedings of the 28th ACM joint meeting on european software
  engineering conference and symposium on the foundations of software
  engineering}, 2020, pp. 1556--1560.

\bibitem{just2014defects4j}
R.~Just, D.~Jalali, and M.~D. Ernst, ``Defects4j: A database of existing faults
  to enable controlled testing studies for java programs,'' in
  \emph{Proceedings of the 2014 international symposium on software testing and
  analysis}, 2014, pp. 437--440.

\end{thebibliography}

\end{document}